\documentclass[reprint,amsmath,amssymb,aps,prl,lengthcheck,floatfix,twocolumn,showpacs]{revtex4-2}

\usepackage{graphicx}
\usepackage{dcolumn}
\usepackage{bm}
\usepackage{easyReview}

\usepackage[normalem]{ulem} 
\newcommand\redout{\bgroup\markoverwith
{\textcolor{red}{\rule[.5ex]{2pt}{0.4pt}}}\ULon}
\usepackage{xcolor}

\begin{document}

\title{}

%
%
%

\begin{abstract}

\end{abstract}
\maketitle

\noindent\textbf{de Koning \emph{et al.} Reply:} In our Letter~\cite{deKoning2023} we performed path-integral ground state (PIGS) calculations of dislocation cores in hcp $^4$He to investigate the presence of off-diagonal long-range order (ODLRO) and assess the existence of a finite condensate fraction. We found that ODLRO is absent in all cases, with the systems displaying no significant difference from the insulating defect-free crystal. This result challenges the superfluid-dislocation-network interpretation for mass-flow observations in $^4$He.  

Boninsegni \emph{et al.}~\cite{Boninsegni2023}, hereafter referred to as BKPPS, contend that our criticism of their work is “unjustified”. In particular, BKPPS argue that our disagreeing results are due to the adopted averaging procedure of the one-body density matrix and "an enhanced local pressure at the dislocation core".  In the following, in addition to refuting BKPPS's contestations, we argue that
the origin of  the difference between our results and those reported by BKPPS is rooted in the application of appropriate boundary conditions for meaningful dislocation simulations. 

BKPPS question our approach with respect to the one-body density matrix (OBDM), $\rho\left(\mathbf{r}_1,\mathbf{r}_2\right)$. It is stated that we ignore its nonuniformity and anisotropy and that our averageing “leads to an enormous suppression of $n_0$,  . . .”.

First, the condensate fraction $n_0$ is a scalar quantity, defined for the system as a whole through the $k\rightarrow 0$ limit of the momentum distribution, which, by definition, is isotropic. In fact, as has been shown in ~\cite{Macia2011}, even when $\rho\left(\mathbf{r}_1,\mathbf{r}_2\right)$ is anisotropic. it is only the isotropic component that persists at large separations. Second, the suggestion that a global cell average leads to an enormous suppression of the dislocation signal is incorrect. As demonstrated previously~\cite{Rota2012b}, PIGS simulations detected a significant condensate fraction ($n_0 \sim 10^{-3}$) for a monovacancy in a cell containing 180 atoms, in which the fraction of atoms neighboring the defected region is $\sim 10^{-2}$. In our simulations this fraction is of the same order of magnitude. Moreover, it is crucial to point out that, at the atomic level, a dislocation core is \emph{not} a 1-dimensional object (BKPPS refer to one-dimensional superfluidity). Instead, it constitutes a truly 3-dimensional tube-like region with a cross-section determined by the core radius, which may span several interatomic spacings. Accordingly, the volume occupied by the defected region represents a non-negligible fraction of the simulation cell and if the dislocation core had contributed to a finite condensate fraction, it would have been detected. 

Subsequently, BKPPS raise the issue of the finite size of the cells and argue that our cells would be “at elevated bulk pressure” compared to the bulk crystal, such that the OBDM for the CS and CE dislocations “are suppressed in comparison with the one for the ideal crystal”. 

These arguments are moot, given that all our results were obtained at the \emph{same} number density of 0.0287 \AA$^{-3}$ for which the superfluidity claims were reported in Refs.~\cite{Boninsegni2007,Soyler2009}. 
Furthermore, the finite-size effects in our calculations are smaller than those in Refs.~\cite{Boninsegni2007,Soyler2009}. First, the computational cells in the latter were smaller and substantially more constrained, imposing periodic boundary conditions only along the dislocation line. As discussed in our Letter~\cite{deKoning2023}, it is well-known~\cite{Gehlen1972,Rao1998} that such a straightjacket-like setup with rigid cylindrical geometries can lead to incorrect core structures. In contrast, our calculations are based on standard dislocation-simulation practice~\cite{Bulatov2006} employing cells that also preserve translational symmetry in the glide direction. A second element, which may further aggravate the reliability of the simulations reported in Refs. ~\cite{Boninsegni2007,Soyler2009}, is the fact that sampling in the latter was performed in the grand-canonical ensemble. Such an approach is unjustified for the investigation of intrinsic dislocation-core structures, given that the addition/removal of matter in an essentially crystalline environment may artificially induce disorder and trigger spurious superfluidity. 

Finally, BKPPS argue that our "treatment of exchange cycles ... is insufficient" and that our visual inspection concluding the absence of long exchange cycles is inadequate to establish whether the ground-state properties starting from a (non-orthogonal) trial wave function have been reached.

The measure of exchange cycles is normally employed as a complement to assess the evolution of a finite $T$ calculation, but it is not directly related to any physical observable.  
Indeed, in the PIGS method all the chains are open, with a trial wave function at the end points~\cite{Rota2012}. If this wave function (choosing, e.g., $\Psi_m = 1$ is sufficient~\cite{Rota2012}) is symmetric under the exchange of two particles, the correct Bose symmetry is guaranteed throughout the entire simulation. It is true that, to improve sampling, we use a swap movement in the middle of the open chain but this is not strictly necessary. Furthermore, the acceptance ratio of such swap movements was negligible (as for the case of the defect-free hcp crystal).

M.K. acknowledges support from CNPq, Fapesp grant no. 2016/23891-6 and the Center for Computing in Engineering \& Sciences - Fapesp/Cepid no. 2013/08293-7. W.C. acknowledges support from the U.S. Department of Energy, Office of Basic Energy Sciences, Division of Materials Sciences and Engineering under Award No. DE-SC0010412. J.B.  acknowledges financial support from the Secretaria d'Universitats i Recerca del Departament d'Empresa i Coneixement de la Generalitat de Catalunya, co-funded by the European Union Regional Development Fund within the ERDF Operational Program of Catalunya (project QuantumCat, Ref.~001-P-001644), and the MINECO (Spain) Grant PID2020-113565GB-C21. C.C. acknowledges financial support from the MINECO (Spain) under the ``Ram\'on y Cajal'' fellowship (RYC2018-024947-I). 
\newline

\noindent Maurice de Koning$^{1,2}$, Wei Cai$^3$, \newline
\noindent Claudio Cazorla$^3$ and Jordi Boronat$^4$\newline
\indent $^1$Instituto de F\'{i}sica Gleb Wataghin, Universidade \newline 
\indent Estadual de Campinas, UNICAMP, 13083-859, \newline 
\indent Campinas, S\~{a}o Paulo, Brazil \newline
\indent $^2$ Center for Computing in Engineering \& Sciences, \newline 
\indent Universidade Estadual de Campinas, UNICAMP, \newline 
\indent 13083-861, Campinas, S\~{a}o Paulo, Brazil\newline
\indent $^3$Department of Mechanical Engineering, \newline
\indent Stanford University, Stanford, CA 94305-4040 \newline
\indent $^4$ Departament de Física, Universitat Politècnica de \newline 
\indent Catalunya, Campus Nord B4-B5, \newline 
\indent 08034 Barcelona, Spain

\bibliographystyle{apsrev4-2}

\begin{thebibliography}{10}%
	\makeatletter
	\providecommand \@ifxundefined [1]{%
		\@ifx{#1\undefined}
	}%
	\providecommand \@ifnum [1]{%
		\ifnum #1\expandafter \@firstoftwo
		\else \expandafter \@secondoftwo
		\fi
	}%
	\providecommand \@ifx [1]{%
		\ifx #1\expandafter \@firstoftwo
		\else \expandafter \@secondoftwo
		\fi
	}%
	\providecommand \natexlab [1]{#1}%
	\providecommand \enquote  [1]{``#1''}%
	\providecommand \bibnamefont  [1]{#1}%
	\providecommand \bibfnamefont [1]{#1}%
	\providecommand \citenamefont [1]{#1}%
	\providecommand \href@noop [0]{\@secondoftwo}%
	\providecommand \href [0]{\begingroup \@sanitize@url \@href}%
	\providecommand \@href[1]{\@@startlink{#1}\@@href}%
	\providecommand \@@href[1]{\endgroup#1\@@endlink}%
	\providecommand \@sanitize@url [0]{\catcode `\\12\catcode `\$12\catcode
		`\&12\catcode `\#12\catcode `\^12\catcode `\_12\catcode `\%12\relax}%
	\providecommand \@@startlink[1]{}%
	\providecommand \@@endlink[0]{}%
	\providecommand \url  [0]{\begingroup\@sanitize@url \@url }%
	\providecommand \@url [1]{\endgroup\@href {#1}{\urlprefix }}%
	\providecommand \urlprefix  [0]{URL }%
	\providecommand \Eprint [0]{\href }%
	\providecommand \doibase [0]{https://doi.org/}%
	\providecommand \selectlanguage [0]{\@gobble}%
	\providecommand \bibinfo  [0]{\@secondoftwo}%
	\providecommand \bibfield  [0]{\@secondoftwo}%
	\providecommand \translation [1]{[#1]}%
	\providecommand \BibitemOpen [0]{}%
	\providecommand \bibitemStop [0]{}%
	\providecommand \bibitemNoStop [0]{.\EOS\space}%
	\providecommand \EOS [0]{\spacefactor3000\relax}%
	\providecommand \BibitemShut  [1]{\csname bibitem#1\endcsname}%
	\let\auto@bib@innerbib\@empty
	\bibitem [{\citenamefont {de~Koning}\ \emph {et~al.}(2023)\citenamefont
		{de~Koning}, \citenamefont {Cai}, \citenamefont {Cazorla},\ and\
		\citenamefont {Boronat}}]{deKoning2023}%
	\BibitemOpen
	\bibfield  {author} {\bibinfo {author} {\bibfnamefont {M.}~\bibnamefont
			{de~Koning}}, \bibinfo {author} {\bibfnamefont {W.}~\bibnamefont {Cai}},
		\bibinfo {author} {\bibfnamefont {C.}~\bibnamefont {Cazorla}},\ and\ \bibinfo
		{author} {\bibfnamefont {J.}~\bibnamefont {Boronat}},\ }\href
	{https://doi.org/10.1103/PhysRevLett.130.016001} {\bibfield  {journal}
		{\bibinfo  {journal} {Phys. Rev. Lett.}\ }\textbf {\bibinfo {volume} {130}},\
		\bibinfo {pages} {016001} (\bibinfo {year} {2023})}\BibitemShut {NoStop}%
	\bibitem [{\citenamefont {Boninsegni}\ \emph {et~al.}(0212)\citenamefont
		{Boninsegni}, \citenamefont {Kuklov}, \citenamefont {Pollet}, \citenamefont
		{Prokof'ev},\ and\ \citenamefont {Svistunov}}]{Boninsegni2023}%
	\BibitemOpen
	\bibfield  {author} {\bibinfo {author} {\bibfnamefont {M.}~\bibnamefont
			{Boninsegni}}, \bibinfo {author} {\bibfnamefont {A.~B.}\ \bibnamefont
			{Kuklov}}, \bibinfo {author} {\bibfnamefont {L.}~\bibnamefont {Pollet}},
		\bibinfo {author} {\bibfnamefont {N.~V.}\ \bibnamefont {Prokof'ev}},\ and\
		\bibinfo {author} {\bibfnamefont {B.~V.}\ \bibnamefont {Svistunov}},\
	}\href@noop {} {\bibfield  {journal} {\bibinfo  {journal} {Phys. Rev. Lett.}\
		} (\bibinfo {year} {2023})}\BibitemShut {NoStop}%
	\bibitem [{\citenamefont {Macia}\ \emph {et~al.}(2011)\citenamefont {Macia},
		\citenamefont {Mazzanti}, \citenamefont {Boronat},\ and\ \citenamefont
		{Zillich}}]{Macia2011}%
	\BibitemOpen
	\bibfield  {author} {\bibinfo {author} {\bibfnamefont {A.}~\bibnamefont
			{Macia}}, \bibinfo {author} {\bibfnamefont {F.}~\bibnamefont {Mazzanti}},
		\bibinfo {author} {\bibfnamefont {J.}~\bibnamefont {Boronat}},\ and\ \bibinfo
		{author} {\bibfnamefont {R.~E.}\ \bibnamefont {Zillich}},\ }\href
	{https://doi.org/10.1103/PhysRevA.84.033625} {\bibfield  {journal} {\bibinfo
			{journal} {Phys. Rev. A}\ }\textbf {\bibinfo {volume} {84}},\ \bibinfo
		{pages} {033625} (\bibinfo {year} {2011})}\BibitemShut {NoStop}%
	\bibitem [{\citenamefont {Rota}\ and\ \citenamefont
		{Boronat}(2012{\natexlab{a}})}]{Rota2012b}%
	\BibitemOpen
	\bibfield  {author} {\bibinfo {author} {\bibfnamefont {R.}~\bibnamefont
			{Rota}}\ and\ \bibinfo {author} {\bibfnamefont {J.}~\bibnamefont {Boronat}},\
	}\href {https://doi.org/10.1103/PhysRevLett.108.045308} {\bibfield  {journal}
		{\bibinfo  {journal} {Phys. Rev. Lett.}\ }\textbf {\bibinfo {volume} {108}},\
		\bibinfo {pages} {045308} (\bibinfo {year} {2012}{\natexlab{a}})}\BibitemShut
	{NoStop}%
	\bibitem [{\citenamefont {Boninsegni}\ \emph {et~al.}(2007)\citenamefont
		{Boninsegni}, \citenamefont {Kuklov}, \citenamefont {Pollet}, \citenamefont
		{Prokof'ev}, \citenamefont {Svistunov},\ and\ \citenamefont
		{Troyer}}]{Boninsegni2007}%
	\BibitemOpen
	\bibfield  {author} {\bibinfo {author} {\bibfnamefont {M.}~\bibnamefont
			{Boninsegni}}, \bibinfo {author} {\bibfnamefont {A.~B.}\ \bibnamefont
			{Kuklov}}, \bibinfo {author} {\bibfnamefont {L.}~\bibnamefont {Pollet}},
		\bibinfo {author} {\bibfnamefont {N.~V.}\ \bibnamefont {Prokof'ev}}, \bibinfo
		{author} {\bibfnamefont {B.~V.}\ \bibnamefont {Svistunov}},\ and\ \bibinfo
		{author} {\bibfnamefont {M.}~\bibnamefont {Troyer}},\ }\href
	{http://link.aps.org/abstract/PRL/v99/e035301} {\bibfield  {journal}
		{\bibinfo  {journal} {Phys. Rev. Lett.}\ }\textbf {\bibinfo {volume} {99}},\
		\bibinfo {pages} {035301} (\bibinfo {year} {2007})}\BibitemShut {NoStop}%
	\bibitem [{\citenamefont {S{\"o}yler}\ \emph {et~al.}(2009)\citenamefont
		{S{\"o}yler}, \citenamefont {Kuklov}, \citenamefont {Pollet}, \citenamefont
		{Prokof'ev},\ and\ \citenamefont {Svistunov}}]{Soyler2009}%
	\BibitemOpen
	\bibfield  {author} {\bibinfo {author} {\bibfnamefont {{\,S}.~G.}\
			\bibnamefont {S{\"o}yler}}, \bibinfo {author} {\bibfnamefont {A.~B.}\
			\bibnamefont {Kuklov}}, \bibinfo {author} {\bibfnamefont {L.}~\bibnamefont
			{Pollet}}, \bibinfo {author} {\bibfnamefont {N.~V.}\ \bibnamefont
			{Prokof'ev}},\ and\ \bibinfo {author} {\bibfnamefont {B.~V.}\ \bibnamefont
			{Svistunov}},\ }\href
	{http://link.aps.org/doi/10.1103/PhysRevLett.103.175301} {\bibfield
		{journal} {\bibinfo  {journal} {Phys. Rev. Lett.}\ }\textbf {\bibinfo
			{volume} {103}},\ \bibinfo {pages} {175301} (\bibinfo {year}
		{2009})}\BibitemShut {NoStop}%
	\bibitem [{\citenamefont {Gehlen}\ \emph {et~al.}(1972)\citenamefont {Gehlen},
		\citenamefont {Hirth}, \citenamefont {Hoagland},\ and\ \citenamefont
		{Kanninen}}]{Gehlen1972}%
	\BibitemOpen
	\bibfield  {author} {\bibinfo {author} {\bibfnamefont {P.~C.}\ \bibnamefont
			{Gehlen}}, \bibinfo {author} {\bibfnamefont {J.~P.}\ \bibnamefont {Hirth}},
		\bibinfo {author} {\bibfnamefont {R.~G.}\ \bibnamefont {Hoagland}},\ and\
		\bibinfo {author} {\bibfnamefont {M.~F.}\ \bibnamefont {Kanninen}},\ }\href
	{https://doi.org/10.1063/1.1660850} {\bibfield  {journal} {\bibinfo
			{journal} {J. Appl. Phys.}\ }\textbf {\bibinfo {volume} {43}},\ \bibinfo
		{pages} {3921} (\bibinfo {year} {1972})}\BibitemShut {NoStop}%
	\bibitem [{\citenamefont {Rao}\ \emph {et~al.}(1998)\citenamefont {Rao},
		\citenamefont {Hernandez}, \citenamefont {Simmons}, \citenamefont
		{Parthasarathy},\ and\ \citenamefont {Woodward}}]{Rao1998}%
	\BibitemOpen
	\bibfield  {author} {\bibinfo {author} {\bibfnamefont {S.}~\bibnamefont
			{Rao}}, \bibinfo {author} {\bibfnamefont {C.}~\bibnamefont {Hernandez}},
		\bibinfo {author} {\bibfnamefont {J.~P.}\ \bibnamefont {Simmons}}, \bibinfo
		{author} {\bibfnamefont {T.~A.}\ \bibnamefont {Parthasarathy}},\ and\
		\bibinfo {author} {\bibfnamefont {C.}~\bibnamefont {Woodward}},\ }\href
	{http://dx.doi.org/10.1080/01418619808214240} {\bibfield  {journal} {\bibinfo
			{journal} {Philos. Mag. A}\ }\textbf {\bibinfo {volume} {77}},\ \bibinfo
		{pages} {231} (\bibinfo {year} {1998})}\BibitemShut {NoStop}%
	\bibitem [{\citenamefont {Bulatov}\ and\ \citenamefont
		{Cai}(2006)}]{Bulatov2006}%
	\BibitemOpen
	\bibfield  {author} {\bibinfo {author} {\bibfnamefont {V.~V.}\ \bibnamefont
			{Bulatov}}\ and\ \bibinfo {author} {\bibfnamefont {W.}~\bibnamefont {Cai}},\
	}\href@noop {} {\emph {\bibinfo {title} {Computer simulations of
				dislocations}}}\ (\bibinfo  {publisher} {Oxford University Press},\ \bibinfo
	{year} {2006})\BibitemShut {NoStop}%
	\bibitem [{\citenamefont {Rota}\ and\ \citenamefont
		{Boronat}(2012{\natexlab{b}})}]{Rota2012}%
	\BibitemOpen
	\bibfield  {author} {\bibinfo {author} {\bibfnamefont {R.}~\bibnamefont
			{Rota}}\ and\ \bibinfo {author} {\bibfnamefont {J.}~\bibnamefont {Boronat}},\
	}\href@noop {} {\bibfield  {journal} {\bibinfo  {journal} {J. Low Temp.
				Phys.}\ }\textbf {\bibinfo {volume} {166}},\ \bibinfo {pages} {21} (\bibinfo
		{year} {2012}{\natexlab{b}})}\BibitemShut {NoStop}%
\end{thebibliography}
%

\end{document}